\documentstyle[preprint,aps]{revtex}
\tighten
\draft
\begin{document}
\widetext
\preprint{CLNS 96/1425, HUTP-96/A043}
\bigskip
\bigskip
\title{Three-Family $SU(5)$ Grand Unification in String Theory}
\medskip
\author{Zurab Kakushadze$^1$\footnote{E-mail: 
zurab@string.harvard.edu} and S.-H. Henry Tye$^2$\footnote{E-mail: 
tye@hepth.cornell.edu}}
\bigskip
\address{$^1$Lyman Laboratory of Physics, Harvard University, Cambridge, 
MA 02138\\
and\\
Department of Physics, Northeastern University, Boston, MA 02115\\
$^2$Newman Laboratory of Nuclear Studies, Cornell University,
Ithaca, NY 14853-5001}
\date{September 2, 1996}
\bigskip
\medskip
\maketitle
\begin{abstract}
{}We present two $3$-family $SU(5)$ grand unified models in the
heterotic string theory. One model has $3$ chiral families and $9$ 
pairs of ${\bf 5} + {\overline {\bf 5}}$ Higgs fields, and an 
asymptotically-free $SU(2) \otimes SU(2)$ hidden sector, 
where the two $SU(2)$s have different matter contents. 
The other model has $6$ left-handed and $3$ right-handed
${\bf 10}$s, $12$ left-handed and $9$ right-handed ${\overline {\bf 5}}$s, 
and an asymptotically-free $SU(3)$ hidden sector. At the string scale,
the gauge couplings $g^2$ of the hidden sectors are three times as big 
as that of $SU(5)$. In addition, both models have an anomalous $U(1)$.
\end{abstract}
\pacs{11.25.Mj, 12.10.Dm, 12.60.Jv}
\narrowtext

{}If the superstring theory is relevant to nature, it must contain the 
standard model of strong and electroweak interactions as part of its low 
energy effective field theory (by low energy, we mean below the string scale).
There are at least three possibilities \cite{rev}. 
The string model contains: \\
$\bullet$ ({\em i}) the standard $SU(3)\otimes SU(2) \otimes U(1)$ model;\\
$\bullet$ ({\em ii}) a so-called "guided" grand unified model, 
{\em e.g.}, flipped $SU(5)$ \cite{flip}; or \\
$\bullet$ ({\em iii}) a grand unified theory (GUT). \\
It is natural to incorporate supersymmetry into the above cases. To have 
dynamical supersymmetry breaking, we also need a hidden sector that
is strongly interacting at some scale above the electroweak scale.
In the four-dimensional heterotic string theory, the total 
rank of the gauge symmetry (with $U(1)$ counted 
as 1) must be less than or equal to 22.
In the first two possibilities \cite{other}, typically there are large hidden sectors.
It then follows that there are numerous choices of the hidden sector, and
detailed dynamical analyses are needed to distinguish one from another. 

{}In the third case, {\em i.e.}, grand unified string theory (GUST),
the situation is somewhat different. To incorporate both chiral fermions and
adjoint (or higher representation) Higgs fields
(needed to break the grand unified gauge 
group to the standard model in the effective field 
theory), the grand unified gauge symmetry must be 
realized with a higher-level current algebra. In the four-dimensional 
heterotic string, the room for gauge symmetry is limited.
Since a higher-level current 
algebra takes up extra room, there is less room for the hidden sector, and 
hence fewer possiblities. Level-2 string models have been extensively 
explored in the literature \cite{two}. So far, all the known level-2 GUSTs
either have an even number of chiral families, or additional exotic chiral matter some of which would remain light after the electroweak breaking. This property simply follows from 
the fact that a ${\bf Z}_2$ orbifold \cite{orb} needed to reach a 
level-$2$ current 
algebra has an even number of fixed points, and this number is closely 
related to the number of chiral families.
Recently, $3$-family grand unified string models 
were constructed \cite{three,kt}, using level-3 current algebras. 
The number of possiblities is very limited. In
Ref \cite{kt}, we give one $E_6$ model and three $SO(10)$ models.
They all have $SU(2)$ as the hidden sector gauge group.
This $SU(2)$ is asymptotically-free,
so that there is a chance of dynamical supersymmetry 
breaking via gaugino condensation\cite{drsw}. Because the gauge coupling 
$g^2$ of this $SU(2)$ is three times bigger than that
of the grand unified gauge group at the string scale, the hidden 
sector does seem to get strong at a scale above the electroweak scale. 
Of course, the viability of these models requires more careful 
analyses, in the framework of string phenomenology\cite{gaugino}.

{}A simple robust way to stabilize the dilaton expectation 
value is to have a semi-simple hidden sector, with more than one gaugino 
condensate\cite{kras}. So it is interesting to ask if one can get a 
larger hidden sector in the $3$-family GUST construction, in particular, a
hidden sector with more than one gaugino condensate.
In a separate paper, we shall give a classification of all $3$-family 
$SO(10)$ and $E_6$ models obtainable from our approach.
Since a level-$3$ current algebra takes up more room than the 
corresponding level-2 
current algebra, the room for the hidden sector in the $3$-family GUSTs is 
severely limited. In our classification, the biggest (and only) 
asymptotically-free hidden sector gauge 
symmetry is $SU(2)$. (There is an $SO(10)$ model with $SU(2)^3$ as its 
hidden sector. None of these $SU(2)$s are asymptotically-free at 
the string scale. However, it is possible that some of their matter fields 
become massive via spontaneous symmetry breaking, so the $SU(2)$s can 
become asymptotically-free below that mass scale.)
So an obvious possibility of having a hidden sector with more than one 
asymptotically-free gauge group in a $3$-family GUST is to consider 
$SU(5)$ grand unification \cite{gg}. 
Since the $SU(5)$ gauge group takes up less room in the heterotic string,
more room may be freed up for the hidden sector. 
Indeed, one can get $3$-family $SU(5)$ GUSTs with larger hidden sectors. 
In this paper, we shall present such a model,
with a $SU(2) \otimes SU(2)$ hidden sector, 
where both $SU(2)$s are asymptotically-free, but they 
have different matter contents. The construction of this model automatically
yields another closely related $3$-family $SU(5)$ model with $SU(3)$ as 
its hidden sector.

{}All the $SU(5)$ models in Ref \cite{kt} can be obtained from 
the spontaneous symmetry breakings of 
the $SO(10)$ models, so the hidden sector remains $SU(2)$.
To obtain a larger hidden sector in a $3$-family $SU(5)$ GUST, we shall 
turn on a Wilson line in the $3$-family $SO(10)$ GUST first presented in 
Ref \cite{three}; this Wilson line 
enhances the gauge symmetry in the hidden sector and at the same time 
breaks the $SO(10)$ down to $SU(5) \otimes U(1)$. Depending on the details
of the construction,
there appear two $3$-family $SU(5)$ models, with gauge symmetries
$SU(3)_1 \otimes SU(5)_3 \otimes U(1)^4$ (the $F1(1)$ model) and 
$SU(2)_1 \otimes SU(2)_1 \otimes SU(5)_3 \otimes U(1)^4$ (the $F2(1)$ model).
Their massless spectra are 
given in Table I. Since their properties are closely related to the 
$3$-family $SO(10)$ GUST (the $T1(1)$ model), its massless spectrum is 
also reproduced in Table I to facilitate comparison.
All the $U(1)$ charges are normalized so that the lowest
allowed value is $\pm 1$, with conformal highest weight $r^2/2$. The 
radius $r$ for each $U(1)$ is given at the bottom of Table I.

{}Besides enhanced hidden sectors, there are two new features that arise 
in these two $SU(5)$ models, in comparison to other $3$-family GUSTs. 
The first new feature is the way the net $3$ chiral families appear.
The structures of chiral families of these two 
$SU(5)$ are quite different from that of the $E_6$, 
$SO(10)$, $SU(5)$ and $SU(6)$ 
models of Refs \cite{three,kt}. 
The second new feature is the presence of an anomalous $U(1)$.
In string theory, this $U(1)$ anomaly is well understood via the 
Green-Schwarz mechanism \cite{gs}.
In contrast, all the known $3$-family $E_6$ and $SO(10)$ models are 
completely anomaly-free. One of the features that these new models have in common with the $SO(10)$ and $E_6$ models is that there is only one adjoint in the grand unified gauge group. Also note that in all of these models the adjoint carries no other quantum numbers \cite{dienes}. 

{}The $SU(5)$ model given in the third column of Table I 
(the $F2(1)$ model) has $3$ left-handed and no right-handed 
${\bf 10}$s, and $12$ left-handed and $9$ right-handed ${\overline {\bf 5}}$s
of $SU(5)$. This means the $F2(1)$ model has $3$ chiral families and 
$9$ pairs of Higgs fields in the fundamental representation.
One of the $U(1)$s, namely, the third one, is anomalous in this model.
The total anomaly is given by $(0,0,+72,0)_L$ (with normalization radius 
$r={1\over {3\sqrt{2}}}$). 
The first two $U(1)$s play the role of the messenger sector, 
while the last $U(1)$ is part of the visible sector. From 
Table I, we see that the two $SU(2)$s in the hidden sector have 
different matter contents, so they are expected to have different 
threshold corrections.
Naively, the first $SU(2)$ in the hidden sector gets 
strong at a scale a little above the electroweak scale, 
while the second $SU(2)$ is still weak at this scale. 
However, this is not expected to happen. Instead,
the anomalous $U(1)$ gauge symmetry is expected to be broken
by the Fayet-Iliopoulos term\cite{fit} at some scale (say, slightly
below the string scale). Presummably a number of scalar fields will develop
vacuum expectation values without breaking supersymmetry. In particular,
there is only one (non-abelian) singlet 
(namely the $({\bf 1},{\bf 1},{\bf 1})(0,0,-6,0)_L$) in the massless 
spectrum of this model that has an anomalous $U(1)$ charge only. 
Its scalar component presumably acquires a vacuum expectation value at this
energy scale for the Higgs mechanism. As a result, a number of 
the $SU(2)$ doublets (but not all) will pick up comparable masses.
Below this mass scale, the two $SU(2)$s now have larger $\beta$ coefficients, 
so both will become strong above the electroweak scale. However, they will 
become strong at different scales, each with its own gaugino 
condensate\cite{gaugino}. A more careful analysis is clearly needed 
to see if supersymmetry breaking happens in a way that satibilizes 
the dilaton expectation value to a reasonable value\cite{kras}.

{}It remains an open question if there are other $3$-family GUSTs
that have hidden sectors with multi-gaugino condensates.
A classification of all $3$-family $SU(5)$ models will be very useful.
$SU(6)$ GUST is another possibility.
In any case, it is clear that the total number of such models will be 
very limited. 

{}The $SU(5)$ model given in the second column of 
Table I (the $F1(1)$ model) has $6$ left-handed 
and $3$ right-handed ${\bf 10}$s, and $12$ left-handed and 
$9$ right-handed ${\overline {\bf 5}}$s of $SU(5)$. 
The first three $U(1)$s of the $F1(1)$ model 
are anomalous, whereas the last $U(1)$, which is part of the visible sector,
is anomaly-free. The total $U(1)$ anomaly is given by 
$(+36,-108,-36,0)_L$ (with their normalization radii given in Table I). 
One can always rotate the charges so that only one $U(1)$ is anomalous.
However, there is no singlet that is charged only under this anomalous $U(1)$. 
As a result, we expect that all three $U(1)$s will be broken. Since the
two anomaly-free $U(1)$s (combinations of the first three $U(1)$s)
play the role of the messenger sector,
the messenger sector scale in this model may be quite high (of
the order of string or GUT scale). This is not necessarily a problem since
the $SU(3)$ hidden sector will then become strong also at a rather 
high scale.

{}Since the construction of 
the $SU(5)$ GUSTs here is based on
the $SO(10)$ GUSTs given in Refs \cite{three,kt}, 
and the approach is very similar, the
discussion presented below shall be self-contained but relatively brief. 
The construction uses the asymmetric orbifold framework\cite{orb}. 
First we shall review the construction of the $SO(10)$ model.
Our starting point is a $N=4$ space-time supersymmetric
Narain model\cite{narain}, which we will refer to as $N0$, 
with the lattice $\Gamma^{6,22}=\Gamma^{2,2} \otimes
\Gamma^{4,4} \otimes \Gamma^{16}$. Here $\Gamma^{2,2} =\{(p_R 
\vert\vert p_L ) \}$ is an even self-dual Lorentzian lattice with
$p_R ,p_L \in {\tilde 
\Gamma}^2$ ($SU(3)$ weight lattice), and $p_L - p_R \in \Gamma^2$
($SU(3)$ root lattice). Similarly, $\Gamma^{4,4} =\{(P_R 
\vert\vert P_L ) \}$ is an even self-dual Lorentzian lattice with
$P_R ,P_L \in {\tilde 
\Gamma}^4$ ($SO(8)$ weight lattice), $P_L - P_R \in \Gamma^4$
($SO(8)$ root lattice). $\Gamma^{16}$ is the ${\mbox{Spin}}(32)/{\bf 
Z}_2$ lattice. This model has $SU(3) \otimes SO(8) \otimes SO(32)$ gauge group.

{}Next we turn on Wilson lines that break the $SO(32)$
subgroup to $SO(10)^3 \otimes SO(2)$:
\begin{equation}
 U_1 =({\vec e}_1/2 \vert\vert {\vec 0})({\bf s}^\prime \vert\vert
 {\bf 0}^\prime )({\bf s}\vert {\bf 0}\vert {\bf 0}
 \vert C)~,
\end{equation}
\begin{equation}
 U_2 =({\vec e}_2/2 \vert\vert {\vec 0})({\bf c}^\prime  \vert\vert
 {\bf 0}^\prime )({\bf 0}\vert {\bf s}\vert {\bf 0}
 \vert C)~.
\end{equation}  
Here we are writing the Wilson lines as shift vectors in the $\Gamma^{6,22}$
lattice. Thus, both $U_1$ and $U_2$ are order two (${\bf Z}_2$) shifts. 
The $\Gamma^{2,2}$ right-moving shifts are given by ${\vec e}_1/2$ and 
${\vec e}_2/2$
(${\vec e}_1$ and ${\vec e}_2$ being the simple roots of $SU(3)$,
while ${\vec 0}$ is the identity (null) weight); the left-moving
$\Gamma^{2,2}$ momenta are not shifted. 
The $\Gamma^{4,4}$ right-moving shifts are given by ${\bf s}^\prime$ and  
${\bf c}^\prime$ (${\bf 0}^\prime$,
${\bf v}^\prime$, ${\bf s}^\prime$ and ${\bf c}^\prime$
are the identity, vector, spinor and conjugate spinor weights of $SO(8)$,
respectively);
the left-moving $\Gamma^{4,4}$ momenta are not shifted. 
The $SO(32)$ shifts are given in the $SO(10)^3 \otimes SO(2)$ basis
(${\bf 0}(0)$, ${\bf v}(V)$, ${\bf s}(S)$ and ${\bf c}(C)$
are identity, vector, spinor and anti-spinor weights of $SO(10)(SO(2))$,
respectively). These Wilson lines break the gauge symmetry down to 
$SU(3) \otimes SO(8) \otimes SO(10)^3 \otimes SO(2)$ in the resulting
$N=4$ Narain model, which was referred to as $N1$. 
All the gauge bosons come from the unshifted sector, 
whereas the shifted sectors give rise to massive states only.

{}Before orbifolding the $N1$ model, we will specify the basis that will be 
used in the following. The right-moving $\Gamma^{2,2}$ momenta
will be represented in the $SU(3)$ basis. The twist corresponding to $2\pi/3$
rotations of these momenta will be denoted by $\theta$.
We will use the $SU(3) \supset SU(2) \otimes U(1)$
basis for the left-moving momenta corresponding to the $\Gamma^{2,2}$ 
sublattice. In this basis
${\bf 1}={\bf 1}(0)+{\bf 2}(3) +{\bf 2}(-3)$,
${\bf 3}({\overline {\bf 3}})={\bf 1}(\mp 2) +{\bf 2}(\pm 1)$.
Here the ireps of $SU(3)$ (identity ${\bf 1}$, triplet ${\bf 3}$ and 
anti-triplet ${\overline {\bf 3}}$) are expressed in terms of the 
irreps of $SU(2)$ (identity ${\bf 1}$ and doublet ${\bf 2}$) and the $U(1)$ 
charges are given in brackets. They are normalized to the radius 
$r=1/\sqrt{6}$; so that the conformal
dimension of a state with $U(1)$ charge $Q$ is $h=(rQ)^2/2$.
 
{}The right-moving $\Gamma^{4,4}$ momenta will be represented in the
$SO(8) \supset SU(3) \otimes U(1)^2$ basis. In this basis, 
the $SO(8)$ momenta have two ${\bf Z_3}$ 
symmetries. The first ${\bf Z}_3$ symmetry is that of
$2\pi/3$ rotations in the $SU(3)$ subgroup of $SO(8)$. The second 
${\bf Z}_3$ symmetry corresponds to a $2\pi/3$ rotation in the $U(1)^2$ plane.
Under this rotation
${\bf v}^\prime  \rightarrow {\bf s}^\prime \rightarrow {\bf c}^\prime
\rightarrow {\bf v}^\prime$, which is the well-known triality of
the $SO(8)$ Dynkin diagram.
In the following, the twist corresponding to the simultaneous 
$2\pi/3$ rotations
in both $SU(3)$ and $U(1)^2$ subgroups will be denoted by $\Theta$.

{}The right-moving $\Gamma^{4,4}$ momenta will be represented in the
$SO(8) \supset SU(2)^4$ basis. In this basis, 
under cyclic permutations of, say, the first three $SU(2)$s, the weights
${\bf v}^\prime$, ${\bf s}^\prime$ and ${\bf c}^\prime$ are permuted. 
This is the same as one of the ${\bf Z}_3$ symmetries of $SO(8)$ 
that we considered in the $SU(3)\otimes U(1)^2$ basis, namely,
the triality symmetry of the $SO(8)$ Dynkin diagram. In the following, we will
denote this outer automorphism of the first three $SU(2)$s by ${\cal P}_2$.
We can conveniently describe the ${\cal P}_2$ outer automorphism as a 
${\bf Z}_3$ twist. Let $\eta_1$, $\eta_2$, $\eta_3$ and 
$\eta_4$ be the real bosons of $SU(2)^4$. 
Next, let $\Sigma =(\eta_1 + \omega^2 \eta_2 + \omega \eta_3 )/ \sqrt{3}$, 
so its complex conjugate $\Sigma^\dagger =(\eta_1 + \omega
\eta_2 + \omega^2 \eta_3 )/ \sqrt{3}$ (here $\omega =\exp(2\pi i /3)$),
and $\rho =(\eta_1 + \eta_2 + \eta_3 )/ \sqrt{3}$. In this
basis, we have one complex boson
$\Sigma$, and two real bosons $\rho$ and $\eta_4$. The outer
automorphism of the first three $SU(2)$s that permutes them is now a ${\bf 
Z}_3$ twist of $\Sigma$($\Sigma^\dagger$) that are eigenvectors with
eigenvalues $\omega$ ($\omega^2$). Meantime, $\rho$ and $\eta_4$ are invariant
under this twist, and therefore unaffected. We also note that 
the $SU(2)$ momenta can be expressed in terms of
a one dimensional lattice $\{n/\sqrt{2}\}$, where odd values of $n$ correspond
to the states in the ${\bf 2}$ irrep, whereas the even values describe the
states in the idenity ${\bf 1}$ of $SU(2)$.

{}Finally, we turn to the $SO(10)^3$ subgroup. To obtain $SO(10)_3$ from
$SO(10)^3$, we must mod out by their outer automorphism. In the following we
will denote this outer automorphism by ${\cal P}_{10}$.
Let the real bosons $\phi^I_p$, $I=1,...,5$, correspond to the
$p^{\mbox{th}}$ $SO(10)$ subgroup, $p=1,2,3$. Then ${\cal P}_{10}$ cyclicly
permutes these real bosons: $\phi^I_1 \rightarrow
\phi^I_2 \rightarrow \phi^I_3 \rightarrow \phi^I_1$. We can define new bosons $\varphi^I \equiv {1\over
\sqrt{3}}(\phi^I_1 +\phi^I_2 +\phi^I_3)$; the other ten real bosons are
complexified via linear combinations $\Phi^I \equiv {1\over
\sqrt{3}}(\phi^I_1 +\omega^2\phi^I_2 +\omega \phi^I_3)$ and
$(\Phi^I)^\dagger \equiv {1\over \sqrt{3}}(\phi^I_1 +\omega \phi^I_2 
+\omega^2 \phi^I_3)$, where $\omega =\exp(2\pi i /3)$.
Under ${\cal P}_{10}$, $\varphi^I$ is invariant, 
while $\Phi^I$ ($(\Phi^I)^\dagger$) are eigenstates with 
eigenvalue $\omega$ ($\omega^2$), {\em i.e.}, ${\cal P}_{10}$ acts as a
${\bf Z}_3$ twist on $\Phi^I$ ($(\Phi^I)^\dagger$).

{}Next we introduce the following ${\bf Z}_3 \otimes {\bf Z}_2$ twist on the
$N1$ model:
\begin{equation}
 T_3=(\theta \vert\vert 0\vert 0)
 (\Theta \vert\vert {\cal P}_2 \vert (-\sqrt{2}/3))
 ({\cal P}_{10} \vert 2/3)~,
\end{equation}
\begin{equation}
 T_2=({\vec 0}
 \vert\vert \sqrt{2}/2\vert 0)(-{\bf 1}\vert\vert 0^3 \vert \sqrt{2}/2)
 (0^{15} \vert 0)~.
\end{equation}
Here $T_3$ is a ${\bf Z}_3$ twist that acts as follows. The right-moving
$\Gamma^{2,2}$ momenta (and the corresponding oscillator excitations)
are ${\bf Z}_3$ twisted by the twist $\theta$. The left-moving $\Gamma^{2,2}$
momenta are untouched. This is an asymmetric orbifold. The right-moving
$\Gamma^{4,4}$ momenta (and the corresponding oscillator excitations)
are ${\bf Z}_3$ twisted by the twist $\Theta$. The corresponding left-movers
are twisted by the outer automorphism ${\cal P}_2$, and the last $SU(2)$ 
real boson $\eta_4$ is shifted by $-\sqrt{2}/3$ (this is an order three shift
as $\sqrt{2}$ is a root of $SU(2)$). Lastly, the three $SO(10)$s are twisted
by their outer automorphism ${\cal P}_{10}$, and the $SO(2)$ momenta are
shifted by $2/3$ (this is an order three shift as well since $2$ is 
in the identity weight
of $SO(2)$). Next, $T_2$ is a ${\bf Z}_2$ twist that acts as follows. The
right-moving $\Gamma^{2,2}$ momenta are untouched, whereas their left-moving
counterparts are shifted by $\sqrt{2}/2$ (Note that this is a ${\bf Z}_2$
shift of the momenta in the $SU(2)$ subgroup of $SU(3)$; the $U(1)$ momenta are
separated from those of $SU(2)$ by a single vertical line).  
$\Gamma^{4,4}$ is asymmetrically twisted: The right-movers are twisted 
by a diagonal ${\bf Z}_2$ twist (${\bf 1}$ 
is a $4\times 4$ identity matrix), whereas the left-movers are shifted 
($\sqrt{2}/2$ is a ${\bf Z}_2$
shift of the momenta in the last $SU(2)$ subgroup of $SO(8)$; it is 
separated from the first three $SU(2)$s by a single vertical line).

{}It is easy to verify that the above $T_3$ and $T_2$ twists are 
compatible with 
the Wilson lines $U_1$ and $U_2$, and that the $N1$ model possesses the
corresponding ${\bf Z}_3 \otimes {\bf Z}_2$ (isomorphic to 
${\bf Z}_6)$ symmetry.   
The resulting model was described in detail in Refs \cite{three,kt}, 
so we will not 
repeat that discussion here. For illustrative purposes, we reproduce 
the massless spectrum of this model in the first column of Table I. 
We shall call this model $T1(1)$, where the the number in the bracket 
refers to the choice of modulus $h=1$ in the moduli space that gives 
the $SO(8)$ in the $N0$ model\cite{kt}.

The $T1(1)$ model has $N=1$ space-time 
supersymmetry, and its gauge group is $SU(2)_1 \otimes SU(2)_3 \otimes
SO(10)_3 \otimes U(1)^3$ (the
subscripts indicate the levels of the corresponding
Kac-Moody algebras). This model is completely anomaly free, and its hidden 
sector is $SU(2)_1$, whereas the observable sector is $S0(10)_3 \otimes
U(1)^2$ (the first and the last $U(1)$s in Table I). $SU(2)_3 \otimes U(1)$
plays the role of the messenger/intermediate sector, or horizontal symmetry.
Note that the net number of the chiral $SO(10)_3$ families in this model 
is $5-2=3$.

{}To obtain a three-family $SU(5)_3$ model
with a larger hidden sector, we first add the following Wilson line
\begin{equation}
 U_3 =T^{\prime}_3= ({\vec 0} \vert\vert 0 \vert \sqrt{{2 \over 3}})
 ({\bf 0}^\prime \vert\vert
 ({\sqrt{2}\over 3})^3 \vert 0)(({1\over 3}{1\over 3}{1\over 3}{1\over 3}
 {2\over 3})^3 \vert 0)
\end{equation}
to the $N1$ model. We will refer to the resulting $N=4$ Narain model as $N6$.
In the unshifted sector of the $N6$ model
the gauge group is broken from $SU(3) \otimes
SO(8) \otimes SO(10)^3 \otimes SO(2)$ down to $SU(3)^2 \otimes SU(5)^3\otimes
U(1)^6$ (Note that $SO(8)$ is broken to $SU(3)\otimes U(1)^2$, and each 
$SO(10)$ is broken to $SU(5)\otimes U(1)$). However, there are additional 
gauge bosons that come from the shifted and inverse shifted sectors. These
are in the irreps $({\bf 3},{\overline{\bf 3}})(2)$ and $({\overline{\bf 3}},
{\bf 3})(-2)$ of $SU(3)\otimes SU(3)\otimes U(1)$ , where the $U(1)$ charge 
(which is normalized to $1/\sqrt{6}$) is given in the parentheses. Thus, the
gauge symmetry of the $N6$ model is $SU(6) \otimes SU(5)^3 \otimes U(1)^5$.
This enhancement of gauge symmetry was made possible by breaking the $SO(10)$
subgroups, so that the resulting $SU(5)$ level-3 models can have enhanced
hidden sectors.

{}Next, we can add the $T_3$ twist to the $N6$ model.  
The resulting model has the gauge symmetry
$SU(4)_1\otimes SU(5)_3 \otimes U(1)^3$ (Note that $SU(6)$ is broken down to
$SU(4)\otimes U(1)^2$, and four of the $U(1)$s in the $N6$ models have been
removed by the $T_3$ twist). The number of chiral families of $SU(5)_3$ in 
this model is $9$ as it is the case for other level-3 models constructed from a
single ${\bf Z}_3$ twist \cite{three,kt}.
Therefore, we add the $T_2$ twist to obtain a model
with the net number of three families. We will refer to the final 
model (obtained
via orbifolding the $N6$ model by the $T_3$ and $T_2$ twists) as $F1(1)$.
The $F1(1)$ model has gauge symmetry
$SU(3)_1 \otimes SU(5)_3 \otimes U(1)^4$. Its massless spectrum is given in 
the second column of Table I.  Note that in the $F1(1)$ model
the $SU(3)$ subgroup arises as a result of the breaking $SU(4) \supset
SU(3) \otimes U(1)$.  The net number of chiral families of 
$SU(5)_3$ is three in the $F1(1)$ model. In this table, all the $U(1)$ 
charges are correlated. For example, by 
$({\bf 1},{\bf 2},{\bf 1})({\pm 1},{\mp 3},+3,0)_L$, we mean 
$({\bf 1},{\bf 2},{\bf 1})(+1,-3,+3,0)_L$ plus 
$({\bf 1},{\bf 2},{\bf 1})(-1,+3,+3,0)_L$.

{}In working out the spectra of the $F1(1)$ model it is useful to
view the Wilson line $U_3$ as a ${\bf Z}_3$ twist $T^{\prime}_3$ that acts
on the $T1(1)$ model, even though this twist consists of shifts only and acts 
on the lattice freely, {\em i.e.}, with no fixed points. In this approach
we are orbifolding the $N1$ lattice by the ${\bf Z}_3 \otimes 
{\bf Z}_2\otimes {\bf Z}^{\prime}_3$ twist generated by $T_3$, $T_2$ and 
$T^{\prime}_3$, respectively. Here we find that there are additional 
possibilities. Indeed, since the orders
of $T_3$ and $T^{\prime}_3$ 
are the same (both of them have order three), 
their respective contributions in the one-loop partition function can
have a non-trivial relative phase between them. Let this phase be $\phi(T_3,
T^\prime_3)$. By this we mean that $3 \phi(T_3,T^\prime_3) =0~(\mbox{mod}~1)$
({\em i.e}, $\phi(T_3,T^\prime_3)$ can be $0,1/3,2/3$), and the 
states that survive the $T_3$ projection in the $T^\prime_3$ 
shifted sector must 
have the $T_3$ phase $\phi(T_3,T^\prime_3)$. Similarly, the states 
that survive the $T_3$ projection in the inverse 
shifted sector $(T^\prime_3)^{-1}$ 
must have the $T_3$ phase $-\phi(T_3,T^\prime_3)$. The string
consistency then requires that the states that survive the $T^\prime_3$ 
projection in the $T_3$ twisted sector must 
have the $T^\prime_3$ phase $-\phi(T_3,T^\prime_3)$. Similarly, the states 
that survive the $T^\prime_3$ projection in the inverse 
twisted sector $(T_3)^{-1}$
must have the $T^\prime$ phase $\phi(T_3,T^\prime_3)$.
The gauge symmetry of the resulting
model depends on the choice of $\phi(T_3,T^\prime_3)$. The models with 
$\phi(T_3,T^\prime_3)=0$ and $\phi(T_3,T^\prime_3)=1/3$ are equivalent.
Note that the model with $\phi(T_3,T^\prime_3)=0$ is precisely the $F1(1)$ 
model. The third choice $\phi(T_3,T^\prime_3)=2/3$ leads to a different
model, which we will refer to as $F2(1)$.
The $F2(1)$ model has gauge symmetry 
$SU(2)_1\otimes SU(2)_1 \otimes SU(5)_3 \otimes U(1)^4$. Its massless 
spectrum is given in the third column of Table I.
Here we point out that another advantage in viewing the $U_3$ Wilson line 
as the $T^\prime_3$ twist is  
that the invariant sublattices and numbers of fixed
points for the $T_3$ and $T_2$ twists remain the same as in the $T1(1)$ model, 
so that working out the spectra of the final models becomes easier.

{}Next we 
translate the above twists $T_3$, $T_2$ and $T^\prime_3$ into the generating
vectors $V_i$ and structure constants $k_{ij}$ of the orbifold construction
rules derived in Ref \cite{kt}. These rules are useful in working
out the spectra of the above models when the book-keeping of various phases 
in the partition function
become non-trivial in such asymmetric orbifolds. Thus, the generating
vectors are given by
\begin{eqnarray}
 &&V_0 =(-{1\over 2} (-{1\over 2}~ 0)^3 \vert\vert 0_r ~ 0_r \vert 0~
 0_r ~0_r \vert 0^{5}~0_r^5 ~0_r) ~,\nonumber \\
 &&V_1 =( 0 (-{1\over 3}~{1\over 3})^3 \vert\vert 0_r~0_r \vert ({2\over 3})~
  0_r ~(-{\sqrt{2}\over 3})_r  \vert 
 ({1\over 3})^5 ~0_r^5 ~({2\over 3})_r )~, \nonumber \\
 &&V_2 =( 0 (0~0)(-{1\over 2}~{1\over 2})^2 \vert\vert ({\sqrt{2}\over 2})_r~0_r
 \vert 0 ~0_r ~({\sqrt{2}\over 2})_r \vert 0^5 ~0_r^5 ~0_r )~,\nonumber \\
 &&V_3 =( 0 (0~0)^3 \vert\vert 0_r~(\sqrt{{2\over 3}})_r\vert 0~
  (\sqrt{2\over 3})_r ~0_r \vert 
 (0^5 ~({1\over \sqrt{3}})_r ({1\over \sqrt{3}})_r ({1\over\sqrt{3}})_r ({1\over\sqrt{3}})_r 
 ({2\over\sqrt{3}})_r ~0_r )~, \nonumber \\
 &&W_1 =( 0 (0~{1\over 2})^3 \vert\vert 0_r~0_r \vert ({1\over 2}) ~0_r~ 
 0_r \vert ({1\over 2})^5 ~0_r^5 ~0_r )~,\nonumber \\
 &&W_2 =( 0 (0~0)(0~{1\over 2})^2 \vert\vert 0_r~0_r \vert 0~
  0_r ~0_r \vert0^5 ~0_r^5 ~0_r )~. \nonumber
\end{eqnarray}
Here $V_1$, $V_2$ and $V_3$ correspond to the $T_3$, $T_2$ and $T^{\prime}_3$
twists, respectively. Note that $W_3$ is a null vector since $V_3$ consists
of shifts only. In the $F1(1)$ and $F2(1)$ models we have chosen $k_{00}=0$ 
for definiteness (the
alternative choice $k_{00}=1/2$ would result in equivalent models with
the space-time chiralities of the states reversed). To preserve $N=1$ 
space-time supersymmetry we must put $k_{20}=1/2$ (the other choice  
$k_{20}=0$ would give models with $N=0$ space-time supersymmetry). Finally,
in $k_{13} = \phi(T_3,T^{\prime}_3)$. Note that 
$k_{13}=0$ for the $F1(1)$ model, and
$k_{13}=2/3$ for the $F2(1)$ model. (As we mentioned earlier, the third choice
$k_{13}=1/3$ results in a model which is equivalent to the $F1(1)$ model.)
The rest of the structure constants are completely fixed.
Here, a remark is in order. In working out the spectra of the $F1(1)$ and
$F2(1)$ models, certain care is needed when using the spectrum generating 
formula of Ref \cite{kt}. In particular, the latter is sensitive to the
assignment of $U(1)$ charges (in $SO(10) \supset SU(5) \otimes 
U(1)$) in the untwisted vs twisted sectors. This manifests itself in a slight
modification of the spectrum generated formula which is required by 
string consistency. This ensures that the states in the final models form
irreps of the final gauge group. Without such a modification, the states
in the final model do {\em not} form irreps of the final gauge group. 
 
{}Let us note the following. The $T1(1)$ model that we have started with is
only one of the three different $SO(10)_3$ models considered in 
Ref \cite{kt}. If we start from the $SO(10)_3$
model given in the second column of Table I in Ref \cite{kt} 
(which we will refer to as $T2(1)$), and add the $T^{\prime}_3$ twist,
we still get the same $F1(1)$ and $F2(1)$ models
given in Table I in this paper. If we start from the third $SO(10)$ model,
given in the third column of Table I in Ref \cite{kt} 
(which we will refer to as the $T3$ model),
and add the $T^{\prime}_3$ twist, we get two 3-family
$SU(6)_3$ models, which we will 
refer to as $S1$ and $S2$. 
The spectra of the $S1$ and $S2$ models
are similar to those of the $F1(1)$ and $F2(1)$ models, respectively. The 
corresponding gauge groups are $SU(3)_1 \otimes SU(6)_3 \otimes U(1)^3$ and
$SU(2)_1 \otimes SU(2)_1 \otimes SU(6)_3 \otimes U(1)^3$. One can get these
spectra from those of the $F1(1)$ and $F2(1)$ 
models via replacing $SU(5)_3 \otimes 
U(1)$ (the last $U(1)$) by $SU(6)_3$.
Under the branching $SU(6)\supset SU(5) \otimes U(1)$,
${\bf 6}={\bf 5}(-1)+{\bf 1}(+5)$ and ${\bf 15}={\bf 5}(+4)+{\bf 10}(-2)$.
Note that the $SU(5)_3\otimes U(1)$ 
matter content in the $F1(1)$ and $F2(1)$ models has the underlying $SU(6)$ 
structure. Similarly, the spectra of the $F1(1)$ and $F2(1)$ 
models can be derived
from the spectra of the $S1$ and $S2$ models by giving the Higgs in the 
adjoint of $SU(6)_3$ a vacuum expectation value that breaks it to 
$SU(5)_3 \otimes U(1)$. 
This should make it clear what the spectra of the $S1$ and $S2$ models 
are. In particular, the $S1$ model has 6 copies of ${\bf 15}$, 3 copies of 
${\overline {\bf 15}}$, 9 copies of ${\overline {\bf 6}}$ and 
3 copies of ${\bf 6}$, while
the $S2$ model has 3 copies of ${\bf 15}$, 12 copies of 
${\overline {\bf 6}}$ and 6 copies of ${\bf 6}$.

{}We would like to thank  Gary Shiu and Yan Vtorov-Karevsky for discussions.
The research of S.-H.H.T. was partially supported by National Science 
Foundation. The work of Z.K. was supported in part by the grant NSF PHY-96-02074, and 
the DOE 1994 OJI award. Z.K. would also like to thank Mr. Albert Yu 
and Mrs. Ribena Yu for financial support.

%============================================================================
%\newpage
%\widetext

%%%%%%%%%%%%%%%%%%%%%%%%%%%%%%%%%%%%%%%%%%%%%%%%%%%%%%%%%%%%%%%%%%%%%%%%%%%%%%%
\begin{table}[t]
\begin{tabular}{|c|l|l|l|} 
%%%%%%%%%%%%%%%%%%%%%%%%%%%%%%%%%%%%%%%%%%%%%%%%%%%%%%%%%%%%%%%%%%%%%%%%%%%%%%%
M  & ~~~~~~~$T1(1)$  &  ~~~~~~~~$F1(1)$  &  ~~~~~~~~~$F2(1)$ \\
   &  &  &  \\
%%%%%%%%%%%%%%%%%%%%%%%%%%%%%%%%%%%%%%%%%%%%%%%%%%%%%%%%%%%%%%%%%%%%%%%%%%%%%%%
  & $SU(2)^2 \otimes SO(10)\otimes U(1)^3$ &
    $SU(3)   \otimes SU(5) \otimes U(1)^4$ &
    $SU(2)^2 \otimes SU(5) \otimes U(1)^4$ \\
   \hline
%%%%%%%%%%%%%%%%%%%%%%%%%%%%%%%%%%%%%%%%%%%%%%%%%%%%%%%%%%%%%%%%%%%%%%%%%%%%%%%
   & $({\bf 1},{\bf 1},{\bf 45})(0,0,0)$ & 
   $({\bf 1},{\bf 24})(0,0,0,0)$
   & $ ({\bf 1},{\bf 1},{\bf 24})(0,0,0,0)$ \\
   & $({\bf 1},{\bf 3},{\bf 1})(0,0,0)$  &  
   $2({\bf 1},{\bf 1})(0,0,0,0)_L$
   & $ 2({\bf 1},{\bf 1},{\bf 1})(0,0,0,0)_L$ \\
 $U$ & $ ({\bf 1},{\bf 1},{\bf 1})(0,-6,0)_L$ & 
     $ ({\bf 1},{\bf 1}) (+6,0,0,0)_L$ 
   & $ ({\bf 1},{\bf 1},{\bf 1})(0,0,-6,0)_L$ \\
   & $2 ({\bf 1},{\bf 4},{\bf 1})(0,+3,0)_L$ & 
     $2 ({\bf 1},{\bf 1})(-3, {\pm 3},{\pm 3},0)_L$ &
     $2 ({\bf 1},{\bf 2},{\bf 1})({\pm 1},{\mp 3},+3,0)_L$ \\
   & $2 ({\bf 1},{\bf 2},{\bf 1})(0,-3,0)_L$ &
     $2 ({\overline {\bf 3}},{\bf 1})(+3,-3,+1,0)_L$ & 
     $  ({\bf 2},{\bf 1}, {\bf 1})({\pm 2},0,+3,0)_L$\\
   & & $({\bf 3},{\bf 1})(0,0,-4,0)_L$ & \\
 \hline
%%%%%%%%%%%%%%%%%%%%%%%%%%%%%%%%%%%%%%%%%%%%%%%%%%%%%%%%%%%%%%%%%%%%%%%%%%%%%%%
%%%%%%%%%%%%%%%%%%%%%%%%%%%%%%%%%%%%%%%%%%%%%%%%%%%%%%%%%%%%%%%%%%%%%%%%%%%%%%%
   & $2 ({\bf 1},{\bf 2},{\bf 16})(0,-{1},-{1})_L$ &
     $6 ({\bf 1},{\bf 10})(+1,-1,-{1},-2)_L$ &
     $3 ({\bf 1},{\bf 1},{\bf 10})(0,0,+2,-2)_L$ \\
   & $2 ({\bf 1},{\bf 2},{\bf 10})(0,-{1},+2)_L$ &
     $6 ({\bf 1},{\bf 5})(+1,-1,-1,+4)_L$ &
     $3 ({\bf 1},{\bf 1},{\bf 5})(0,0,+2,+{4})_L$ \\
  $T3$ & $2 ({\bf 1},{\bf 2},{\bf 1})(0,-{1},-{4})_L$ &
     $6 ({\bf 1},{\bf 1})(-2,+1,-1,-{5})_L$ & 
     $6 ({\bf 1},{\bf 1},{\bf 1})(\pm 1,\mp {1},{-1},-5)_L$ \\
   & $ ({\bf 1},{\bf 1},{\bf 16})(0,+{2},-{1})_L$ &
     $6 ({\bf 1},{\overline {\bf 5}})(-2,+1,-{1},+1)_L$ & 
     $6 ({\bf 1},{\bf 1},{\overline {\bf 5}})(\pm 1,\mp {1},-1,+{1})_L$ \\
   & $ ({\bf 1},{\bf 1},{\bf 10})(0,+{2},+{2})_L$ &
     $3 ({\bf 1},{\bf 1})(+1,0,+2,-{5})_L$ &
      \\
   & $({\bf 1},{\bf 1},{\bf 1})(0,+{2},-{4})_L$ &
      $3 ({\bf 1},{\overline {\bf 5}})(+1,0,+2,+1)_L$ &
      \\ \hline
%%%%%%%%%%%%%%%%%%%%%%%%%%%%%%%%%%%%%%%%%%%%%%%%%%%%%%%%%%%%%%%%%%%%%%%%%%%%%%%
  &  &$3({\bf 1},{\bf 1})(+2,+1,+1,+5)_L$
      &
      $ 3({\bf 1},{\bf 1},{\bf 1}) (\pm 1,\pm {1},+1,+{5})_L$ \\
 $T6$  &$ ({\bf 1},{\bf 1},{\overline {\bf 16}}) (\pm 1,+{1},+{1})_L$ &
      $3({\bf 1},{\bf 5})(+2,+1,+1,-1)_L$ &
      $ 3({\bf 1},{\bf 1}, {\bf 5})(\pm 1,\pm {1},+1,-1)_L$ \\
   & $ ({\bf 1},{\bf 1},{\bf 10})(\pm 1,+{1},-{2})_L$  &
    $ 3({\bf 1},{\overline {\bf 10}}) ( -1,-{1},+{1},+2)_L$
       &
      \\
   &$({\bf 1},{\bf 1},{\bf 1})(\pm 1,+{1},+{4})_L$ &  
 $ 3({\bf 1},{\overline {\bf 5}})( -1,-{1},+1, -4)_L$
    & \\
 \hline 
%%%%%%%%%%%%%%%%%%%%%%%%%%%%%%%%%%%%%%%%%%%%%%%%%%%%%%%%%%%%%%%%%%%%%%%%%%%%%%%
   &  $({\bf 2},{\bf 2},{\bf 1})(0,0,0)_L$ &
      $({\bf 3},{\bf 1})(\pm 3,-3,-1,0)_L$ &
  $({\bf 2},{\bf 2},{\bf 1})(\pm 1,\mp 3,0,0)_L$ \\
 $T2$ &  $({\bf 2},{\bf 4},{\bf 1})(0,0,0)_L$ &
       $({\overline {\bf 3}},{\bf 1})(-3,+3,+1,0)_L$ &
 $({\bf 1},{\bf 2},{\bf 1})(\pm 1,\pm 3, -3,0)_L$ \\
   & $({\bf 1},{\bf 1},{\bf 1})(\pm 3,-3,0)_L$ &
   $ ({\bf 1},{\bf 1})(+3,\pm 3,\mp 3,0)_L$ & \\
 \hline
%%%%%%%%%%%%%%%%%%%%%%%%%%%%%%%%%%%%%%%%%%%%%%%%%%%%%%%%%%%%%%%%%%%%%%%%%%%%%%%
 $U(1)$ & ~~~~$({1\over \sqrt{6}},~{1\over {3\sqrt{2}}},~{1\over 6})$ &
    $({1\over{3\sqrt{2}}},~{1\over{2\sqrt{3}}},~{1\over 
    {2\sqrt{3}}},~{1\over {3\sqrt{10}}})$ &
  $({1\over 2},~{1\over {2\sqrt{3}}},~{1\over {3\sqrt{2}}},~{1\over
  {3\sqrt{10}}})$~ \\
%%%%%%%%%%%%%%%%%%%%%%%%%%%%%%%%%%%%%%%%%%%%%%%%%%%%%%%%%%%%%%%%%%%%%%%%%%%%%%%
\end{tabular}
%%%%%%%%%%%%%%%%%%%%%%%%%%%%%%%%%%%%%%%%%%%%%%%%%%%%%%%%%%%%%%%%%%%%%%%%%%%%%%%
\caption{The massless spectra of the three models: $T1(1)$, $F1(1)$ and 
$F2(1)$, with gauge symmetries : 
(1) $SU(2)_1\otimes SU(2)_3 \otimes SO(10)_3\otimes U(1)^3$, 
(2) $SU(3)_1 \otimes SU(5)_3 \otimes U(1)^4$, and 
(3) $SU(2)_1 \otimes SU(2)_1 \otimes SU(5)_3 \otimes U(1)^4$. 
Note that double signs (as in $({\bf 1},{\bf 2},{\bf 1})(\pm 1,\mp 3, -3,0)_L$)
are correlated. The $U(1)$ normalization 
radii are given at the bottom of the table.
The graviton, dilaton and gauge supermultiplets are not shown.}

\end{table}
%%%%%%%%%%%%%%%%%%%%%%%%%%%%%%%%%%%%%%%%%%%%%%%%%%%%%%%%%%%%%%%%%%%%%%%%%%%%%%%


\begin{references}
\bibitem{rev} For recent reviews, see,  {\em e.g.},\\
A.E. Faraggi, preprint IASSNS-HEP-94/31 (1994), hep-ph/9405357; \\ 
L.E. Ib\'a\~nez, preprint FTUAM-95-15 (1995), hep-th/9505098; \\
C. Kounnas, preprint CERN-TH/95-293, LPTENS-95/48 (1995), hep-th/9512034;\\
Z. Kakushadze and S.-H.H. Tye, in Proceedings of the International Symposium on Heavy Flavor and Electroweak Theory: August 16-19, 1995, Beijing, China / edited by C.-H. Chang and C.-S. Huang (World Scientific, Singapore, 1996) p. 264, hep-th/9512155; \\
J. Lopez, Rept. Prog. Phys. {\bf 59} (1996) 819, hep-ph/9601208; \\
K.R. Dienes, preprint IASSNS-HEP-95/97 (1996), hep-th/9602045; \\ 
F. Quevedo, CERN-TH/96-65 (1996), hep-th/9603074; \\
J.D. Lykken, hep-th/9607144.

\bibitem{flip} I. Antoniadis, J. Ellis, J. Hagelin and D.V. Nanopoulos (Ref \cite{other}).

\bibitem{other} For a partial list, see, {\em  e.g.},\\
L. E. Ib\'a\~nez, J.E. Kim, H. P. Nilles and F. Quevedo,
Phys. Lett. {\bf B191} (1987) 282; \\
L. E. Ib\'a\~nez, H. P. Nilles and F. Quevedo, Nucl. Phys. {\bf B307}
(1988) 109; \\
A. Font, L. E. Ib\'a\~nez, and F. Quevedo,
Nucl. Phys. {\bf B345} (1990) 389;\\
I. Antoniadis, J. Ellis, J. Hagelin and D.V. Nanopoulos,
Phys. Lett. {\bf B194} (1987) 231; {\bf B208} (1988) 209; 
{\bf B231} (1989) 65;\\
J. Lopez, D.V. Nanopoulos and K. Yuan, Nucl. Phys. {\bf B399} (1993) 654;\\
J. Lopez and D.V. Nanopoulos,  Nucl. Phys. {\bf B338} (1989) 73; Phys. Rev.
Lett. {\bf 76} (1996) 1566;\\
I. Antoniadis, G.K. Leontaris and J. Rizos, Phys. Lett. {\bf B245} 
(1990) 161;\\
G.K. Leontaris, Phys. Lett. {\bf B372} (1996) 212;\\
D. Finnell, Phys. Rev. {\bf D53} (1996) 5781;\\
A. Maslikov, S. Sergeev and G. Volkov, Phys. Rev. {\bf D50} (1994) 7440;\\
A. Maslikov, I. Naumov and G. Volkov, Int. J. Mod. Phys. {\bf A11} (1996) 
1117;\\
S. Chaudhuri, G. Hockney, J.D. Lykken, Nucl. Phys. {\bf B469} 357;\\
D. Bailin, A. Love and S. Thomas, Phys. Lett. {\bf B194} (1987) 385;\\
B.R. Greene, K.H. Kirklin, P.J. Miron and G.G. Ross, Nucl. Phys. 
{\bf B292} (1987) 606;\\
R. Arnowitt and P. Nath, Phys. Rev. {\bf D40} (1989) 191;\\
A. Font, L.E. Ib\'a\~nez, F. Quevedo and A. Sierra,
Nucl. Phys. {\bf B331} (1990) 421;\\ 
A.E. Faraggi, Phys. Lett. {\bf B278} 
(1992) 131; Phys. Rev. {\bf D47} (1993) 5021;\\  
L.E. Ib\'a\~nez, D. L\"{u}st and G. G. Ross, Phys. Lett. {\bf B272} 
(1991) 251;\\ 
S. Kachru, Phys. Lett. {\bf B349} (1995) 76.

\bibitem{two} See, {\em e.g.},\\
D. C. Lewellen, Nucl. Phys. {\bf B337} (1990) 61; \\
J. A. Schwartz, Phys. Rev. {\bf D42} (1990) 1777; \\
J. Ellis, J. Lopez and D.V. Nanopoulos, Phys. Lett. 
{\bf B245} (1990) 375; \\
S. Chaudhuri, S.-W. Chung, G. Hockney and J.D. Lykken,
Nucl. Phys. {\bf B456} (1995) 89; \\
G. B. Cleaver, Nucl. Phys. {\bf B456} (1995) 219; \\
G. Aldazabal, A. Font, L.E. Ib\'a\~nez and A. M. Uranga,
Nucl. Phys. {\bf B452} (1995) 3; \\
J. Erler, Nucl. Phys. {\bf B475} (1996) 597;\\
Z. Kakushadze, G. Shiu and S.-H.H. Tye, CLNS 96/1414, hep-th/9607137.

\bibitem{orb}L. Dixon, J. Harvey, C. Vafa and E. Witten, Nucl. Phys.
{\bf B261} (1985) 620; {\bf B274} (1986) 285; \\
K.S. Narain, M.H. Sarmadi and C. Vafa, Nucl. Phys.
{\bf B288} (1987) 551; \\
L.E. Ib\'a\~nez, H.P. Nilles and F. Quevedo, Phys. Lett. {\bf 187B} (1987) 25.

\bibitem{three} Z. Kakushadze and S.-H.H. Tye, Phys. Rev. Lett. {\bf 77}
(1996) 2612, hep-th/9605221.

\bibitem{kt} Z. Kakushadze and S.-H.H. Tye, preprint CLNS 96/1413
(1996), hep-th/9607138.

\bibitem{drsw}J.-P. Derendinger, L.E. Ib\'a\~nez and H.P. Nilles, Phys. Lett.
{\bf 155B} (1985) 65; \\
M. Dine, R. Rohm, N. Seiberg and E. Witten, Phys. Lett.
{\bf 156B} (1985) 55.

\bibitem{gaugino}On superstring phenomenology, see, {\em e.g.},\\
T. Taylor, G. Veneziano and S. Yankielowicz, Nucl. Phys. {\bf B218} 
(1983) 493; \\
T. Taylor, Phys. Lett. {\bf 164B} (1985) 43; \\
A. Font, L.E. Ib\'a\~nez, D. L\"ust and F. Quevedo,
Phys. Lett. {\bf 245B} (1990) 401; \\
S. Ferrara, N. Magnoli, T. Taylor and G. Veneziano, Phys. Lett. {\bf 245B}
(1990) 409; \\
H.P. Nilles and M. Olechowski, Phys. Lett. {\bf 248B} (1990) 268; \\
P. Bin\'etruy and M.K. Gaillard, Phys. Lett. {\bf 253B} (1991) 119; \\
L. Dixon, V. Kaplunovsky and J. Louis, Nucl. Phys. {\bf B329} (1990) 27; 
{\bf B355} (1991) 649; \\
V. Kaplunovsky, Nucl. Phys. {\bf B307} (1988) 145; {\bf 382} (1992) 436; \\
V. Kaplunovsky and J. Louis, Nucl. Phys. {\bf B444} (1995) 191.

\bibitem{kras}N.V. Krasnikov, Phys. Lett. {\bf 193B} (1987) 37; \\
L. Dixon, V. Kaplunovsky, J. Louis and M. Peskin, SLAC-PUB-5229 (1990); \\
J.A. Casas, Z. Lalak, C. Mu\~noz and G.G. Ross, Nucl. Phys, {\bf B347} (1990)
243; \\
T. Taylor, Phys. Lett. {\bf B252} (1990) 59.

\bibitem{gg} H. Georgi and S.L. Glashow, Phys. Rev. Lett. {\bf 32} (1974) 438.

\bibitem{gs} M. Green and J.H. Schwarz, Phys. Lett. {\bf 149B} (1984) 117; \\
E. Witten,  Phys. Lett. {\bf 149B} (1984) 351.

\bibitem{dienes}K.R. Dienes and J. March-Russell, Nucl. Phys. {\bf B479} (1996) 113; \\
K.R. Dienes, IASSNS-HEP-96/64 (1996), hep-ph/9606467. 

\bibitem{fit}M. Dine, N. Seiberg and E. Witten, Nucl. Phys. {\bf B289}
(1987) 585; \\
J. Attick, L. Dixon and A. Sen, Nucl. Phys. {\bf B292} (1987) 109; \\
M. Dine, I. Ichinose and N. Seiberg, Nucl. Phys. {\bf B293} (1987)253.

\bibitem{narain} K.S. Narain, Phys. Lett. {\bf 169B} (1986) 41; \\
K.S. Narain, M.H. Sarmadi and E. Witten, Nucl. Phys. {\bf B279} (1987) 369.

\end{references}
\end{document}